\documentclass[aps,pra,twocolumn,notitlepage,superscriptaddress,showkeys,longbibliography]{revtex4-1}

\usepackage{natbib}
\usepackage{amsmath, amsthm, amssymb}
\usepackage{graphicx}
\usepackage{xcolor}
\usepackage{xargs}
\usepackage[colorlinks,linkcolor=blue,hyperindex,CJKbookmarks]{hyperref}
\usepackage[utf8]{inputenc}
\usepackage{ragged2e}

\begin{document}

\title{Negative $c$-axis longitudinal magnetoresistance in FeSe}

\author{M. Lourdes Amig\'o}
\affiliation{Centro At\'omico Bariloche (CNEA), CONICET, 8400 Bariloche, R\'io Negro, Argentina.}

\author{Jorge I. Facio}
\affiliation{Centro At\'omico Bariloche (CNEA), CONICET, 8400 Bariloche, R\'io Negro, Argentina.}
\affiliation{Instituto Balseiro (U. N. Cuyo), R\'io Negro, Argentina.}

\author{Gladys Nieva} \email{gladys.nieva@ib.edu.ar}
\affiliation{Centro At\'omico Bariloche (CNEA), CONICET, 8400 Bariloche, R\'io Negro, Argentina.}
\affiliation{Instituto Balseiro (U. N. Cuyo), R\'io Negro, Argentina.} 

\date{\today}

\begin{abstract}
Below the structural transition occurring at $T_s=90$\,K, FeSe exhibits positive transverse magnetoresistance when the current is applied parallel to the $ab$-plane. In this study, we show that, in contrast, when both the magnetic field and the current are aligned along the $c$-axis, the magnetotransport changes significantly. In this configuration, FeSe develops a sizable negative longitudinal magnetoresistance ($\sim$15\% at $T$=10\,K and $\mu_0H$=16\,T) in the nematic phase. We attribute this finding to the effect of the applied magnetic field on the scattering from spin fluctuations. Our observations reflect the intricate interplay between spin and orbital degrees of freedom in the nematic phase of FeSe.
\end{abstract}

\maketitle

\section{Introduction}\label{Introduction}

In recent years, the normal state of the iron-based superconductors has been widely studied to understand the origin of its superconducting state \cite{fernandes2014drives,vaknin2016magnetic,kreisel2020remarkable,bohmer2022nematicity,rhodes2022fese}. 
 A common characteristic shared by many members of this family is a structural transition from tetragonal to orthorhombic symmetry at a temperature $T_s$. The orthorhombic state is called nematic, indicating that the underlying rotational symmetry of the electronic state of the high-temperature phase has been reduced. 
 Early theoretical \cite{PhysRevB.77.224509} and experimental \cite{chu2010plane,PhysRevB.81.184508,Yi6878} works in pnictides established that this transition is driven by electronic spin or orbital degrees of freedom, rather than by phonons.
In pnictides, the transition occurs simultaneously with or precedes long-range magnetic order, suggesting that spin fluctuations is the driving force of the nematic order \cite{bohmer2022nematicity, fernandes2022iron}.
On the other hand, FeSe does not present long-range magnetic order \cite{PhysRevLett.103.057002}, making the question on the origin of its nematicity more complex. 

In the tetragonal phase, FeSe presents a Fermi surface consisting of nearly cylindrical hole and electron sheets centered at the $\Gamma$ and M points of the Brillouin zone, respectively \cite{watson2015emergence}. 
The bands display a small but observable dispersion along the $k_z$ momentum direction, which reaches approximately 10 meV for the top of the hole bands.
At the structural transition ($T_s$=90\,K), ARPES experiments observe a significant band structure reconstruction \cite{shimojima2014lifting, nakayama2014reconstruction,PhysRevB.94.155138,PhysRevX.9.041049}, supporting the notion that the driving force involves electronic degrees of freedom \cite{fedorov2016effect}. 
 However, since orbital and spin degrees of freedom are naturally coupled, distinguishing the primary order parameter --whether it corresponds to orbital order or is associated with spin order-- remains challenging \cite{fernandes2014drives}. In this context, several scenarios have been considered to understand the origin of the structural transition and its potential implications for the superconducting phase \cite{massat2016charge,baek2015orbital,PhysRevLett.114.027001,PhysRevX.6.021032,PhysRevLett.116.227001,wang2015strong,wang2015nematicity,glasbrenner2015effect,wang2016magnetic,benfatto2018nematic}.

In addition, several quantities of interest exhibit profound changes at $T_s$.
The multiband character of FeSe becomes more evident in transport properties below $T_s$ due to the lifting of $xz$/$yz$ orbital degeneracy. 
Positive transverse magnetoresistance in the $ab$-plane appears below the transition and has been related to the reconstruction of the bands at $T_s$ \cite{kasahara2014field,PhysRevB.90.144516,amigo2014multiband,PhysRevLett.115.027006,knoner2015resistivity}. Changes in the Hall resistivity have also been associated with variations in the number and mobility of the carriers \cite{PhysRevLett.115.027006}.
The in-plane resistivity develops a sizable anisotropy between the $a$ and $b$ axis \cite{PhysRevLett.117.127001,PhysRevX.11.021038}, and at low temperature, the nematic phase exhibits complex high-field magnetrotransport properties \cite{kasahara2016giant,PhysRevResearch.2.013309}.
Moreover, even though long-range magnetic order is absent in FeSe, short-range anisotropic magnetic fluctuations have been reported to increase below $T_s$ \cite{baek2015orbital,PhysRevLett.114.027001,wang2015strong} and the strong spatial anisotropy of these fluctuations has been associated with the spin-orbit coupling \cite{ma2017prominent}.

In this work, we focus on the $c$-axis transport, an aspect of FeSe that has been relatively less studied, with previous works primarily addressing properties of the superconducting phase \cite{sadakov2015c,PhysRevB.93.064503,sinchenko2017gossamer}.
Here, we study the magnetotransport properties across the nematic transition and uncover that when the current and the magnetic field are along the $c$-axis,  FeSe develops \textit{negative} longitudinal magnetoresistance (NLMR) below $T_s$.  
We argue that the NLMR originates in the enhancement of short-range anisotropic magnetic fluctuations that occurs below $T_s$. Our results provide new insights into the relevance of spin fluctuations  in the nematic phase of FeSe.

The remainder of this article is organized as follows. Section \ref{methods} presents the experimental methods related to crystal growth and transport measurements. Section \ref{results} describes the transport measurement results and Section \ref{discussion} presents our interpretation of the experimental data. Section \ref{conclusions} contains  our concluding remarks, and  Appendix \ref{ApenA} and \ref{ApenB} provide additional measurements on various samples.

\section{Methods}
\label{methods}

Single crystals of FeSe were grown in a flux of AlCl$_3$/KCl \cite{amigo2014multiband, amigo2017vortex}. X-ray diffraction and Energy-dispersive x-ray spectroscopy (EDX) were used to characterize the crystal structure and the composition of the as-grown samples. The crystals have a platelet morphology and the only phase present is identified as the tetragonal
phase of FeSe at room temperature.

The resistivity and magnetoresistivity measurements were conducted in an OXFORD cryostat equipped with a superconducting magnet with a maximum magnetic field of 16\,T. The crystals were mounted on a rotatable sample holder capable of controlling the angle with an angular resolution of 0.05\,$^\circ$.

For transport measurements with the current in the $ab$-plane ($I$$\parallel$$ab$), we employed the conventional four wire method. Meanwhile, for measurements with the current along the $c$-axis ($I$$\parallel$$c$), we implemented a modified Corbino configuration, with perfectly aligned contacts patterned on the top and bottom of the sample, as illustrated in the sketch of figure \ref{rvst}(b). In both cases, we used sputtered gold patterns and gold wires glued with silver paint or silver epoxy. The results presented in the main text of this paper were obtained with the same crystal for the measurements with $I$$\parallel$$ab$ or with $I$$\parallel$$c$. When we changed the contact configuration, the surface was cleaved to remove any traces of the gold pattern or silver contacts. 

The most challenging aspect of the measurements presented here is the correct alignment of the contacts in the modified Corbino configuration used to measure the $c$-axis transport.
As detailed below, the presence of a maximum in the resistivity, occurring at different temperatures for the $ab$-plane and $c$-axis resistivities, helps address this issue.
To verify the validity of the methodology, Appendix \ref{ApenA} presents $c$-axis transport measurements for different crystals. 

Lastly, we have also measured the Hall resistivity within the $ab$-plane. As shown in Appendix \ref{ApenB}, the temperature dependency of the Hall coefficient of our samples is in agreement with previous reports \cite{kasahara2014field, PhysRevLett.115.027006}.

\section{Results}
\label{results}

\begin{figure*}[t]
\begin{center}
\includegraphics[width=0.9\textwidth]{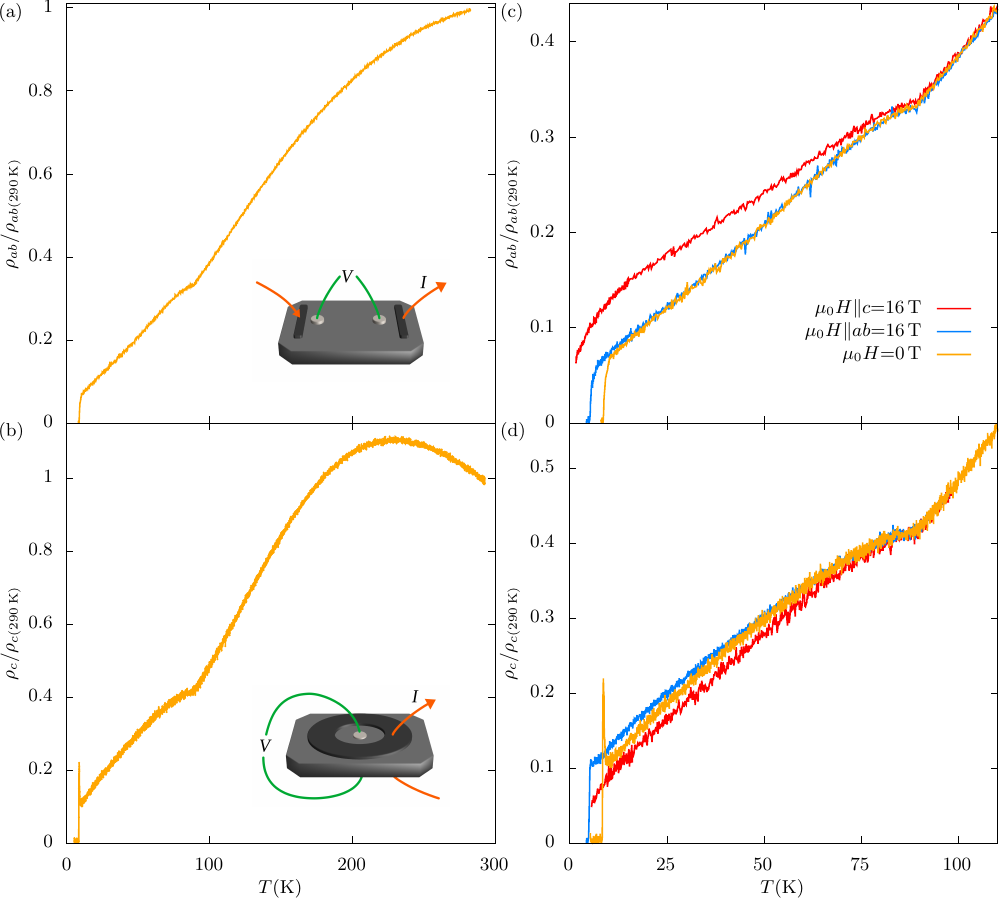}
	\caption{Temperature dependence of the normalized resistivity with the current applied (a) in the $ab$-plane and (b) along the $c$-axis for $\mu_0 H$=0\,T, with a sketch of the contacts configuration presented as an inset in each figure. Normalized resistivity for $\mu_0 H$=0 and 16\,T, parallel (in blue) or perpendicular (in red) to the $ab$-plane, with the current (c) in the $ab$-plane or (d) along the $c$-axis.}
\label{rvst}
\end{center}
\end{figure*}

Figures \ref{rvst}(a) and \ref{rvst}(b) present the normalized resistivity for a single crystal of FeSe with the current parallel to the $ab$-plane ($\rho_{ab}$) or along the $c$-axis ($\rho_c$). In both configurations, the resistivity drops to zero at the superconducting transition, with $T_c$=8.7(2)\,K.
For the current along the $c$-axis, the resistivity shows a sharp upturn at the onset of the superconducting transition.
In other superconductors, this behavior has been associated with how the supercurrent percolates between the electrical contacts in the presence of  small inhomogeneities in the $T_c$ \cite{PhysRevLett.66.2254,PhysRevB.47.15302,MOSQUEIRA19941397,10.1063/1.3696888}.
It is likely this effect becomes observable in the $c$-axis resistivity due to the small extension of the samples of FeSe along this direction.

 The change in the slope at $T_s$=90(1)\,K is related to the structural transition from tetragonal to orthorhombic symmetry at low temperatures. 
 Approaching room temperature, FeSe enters a regime characterized by a saturation of the resistivity and the presence of a resistivity maximum occurring at different temperatures for $I$$\parallel$$ab$ and $I$$\parallel$$c$, specifically $T_{ab}^{max}$=315(3)\,K  and $T_c^{max}$=229(1)\,K, respectively.
 The former value is determined by extrapolating the derivative of $\rho_{ab}$ with respect to the temperature to zero.
The maximum in the $ab$-plane resistivity has been previously reported in the literature \cite{0953-2048-28-10-105009, abdel2016impurity} 
  while the one along the $c$-axis can be seen in \cite{sadakov2015c}.

For temperatures higher than $T_s$, the resistivity is nearly independent of the direction and magnitude of the applied magnetic field for both current directions, with the magnetoresistivity below 1\%, as shown in figures \ref{rvst}(c) and  \ref{rvst}(d). 
Below the structural transition, the system exhibits anisotropic magnetoresistance. 
For $I$$\parallel$$ab$, the observed behavior is well-established in FeSe:  the transverse magnetoresistance is positive for $H \mathord \parallel c$ and almost negligible for $H \mathord \parallel ab$ \cite{kasahara2014field,PhysRevB.90.144516,amigo2014multiband,PhysRevLett.115.027006,knoner2015resistivity}. 
This phenomenology can be understood based on multiband models (see, e.g., \cite{amigo2014multiband}), where the emergence of magnetoresistance below $T_s$ is related to the electronic structure reconstruction observed in several ARPES experiments \cite{shimojima2014lifting, nakayama2014reconstruction, fedorov2016effect, watson2016evidence}.

Conversely, when the current flows along the $c$-axis, the behavior of the magnetoresistance differs markedly, as presented in figure \ref{rvst}(d). 
For $H \mathord \parallel ab$, the transverse magnetoresistance is positive, yet it turns negative in the longitudinal case, $H \mathord \parallel c$. To the best of our knowledge, this is the first report of NLMR in FeSe. 
Similar to the case with $I$$\parallel$$ab$, the pronounced increase in magnetoresistance upon cooling below $T_s$ suggests that the NLMR is intimately connected to the nematic and structural transitions.

\begin{figure}[t]
\begin{center}
\includegraphics[width=0.48\textwidth]{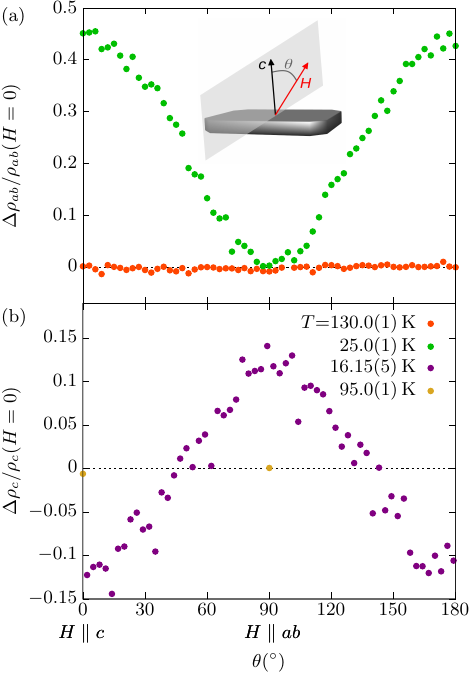}
\end{center}
	\caption{a) Angular dependence of the resistivity with the current in the $ab$-plane under a magnetic field for $\mu_0 H$=16\,T. b) Analogous data for the current along the $c$-axis. In both panels, the dotted line represents the zero field resistivity. }  
	\label{rvsth}
\end{figure}

Figure \ref{rvsth} presents the magnetoresistivity as a function of the angle, $\theta$, between $H$ and the $c$-axis.  For both current configurations, the field rotation occurs within the same plane.
A sketch of the rotation of the magnetic field is shown as an inset in the same figure. 
The angular dependence of the resistivity is determined by the anisotropy of the magnetoresistance, resulting in no variation for temperatures above $T_s$.
Figure \ref{rvsth}(a) corresponds to the configuration with $I$$\parallel$$ab$. In this setup, the magnetic field and the current are always perpendicular to each other, and the magnetoresistance remains equal or above zero for all values of $\theta$. 
Figure \ref{rvsth}(b) shows the case with $I$$\parallel$$c$, where the magnetic field orientation ranges from parallel to the current ($\theta$=0\,$^\circ$) to perpendicular ($\theta$=90\,$^\circ$). 
The behavior is markedly different from that with the current in the $ab$-plane. Notably,  for a wide angular range, the magnetoresistance is negative. 
This observation indicates that a possible misalignment between the current and the magnetic field is not critical for observing negative magnetoresistance.

On the other hand, the correct alignment of the voltage contacts is essential. If not achieved, the measurement would effectively probe a mixture of the resistivities of the $c$-axis and the $ab$-plane. Since the magnetoresistance of the latter is positive, this could lead to an overall positive magnetoresistance.
In this regard, we find that in FeSe the evolution of $\rho_c$ with temperature for $H$=0 provides an independent test for the contacts alignment. Specifically, a slight misalignment shifts the maximum $T_{c}^{max}$ to higher temperatures because $T_{ab}^{max} > T_{c}^{max}$ (see figures \ref{rvst}(a) and \ref{rvst}(b)).
In the several samples measured, we have observed that even a misalignment causing an increase of $T_{c}^{max}$ of only 5\,K results in the disappearance of the negative magnetoresistance, and the angular dependence takes on the shape typical for the current configuration in the $ab$-plane. This observation could very well explain why the NLMR has so far remained elusive. Appendix \ref{ApenA} demonstrates the repeatability of the NLMR and its angular dependence for more crystals of FeSe.

\begin{figure}[t]
\begin{center}
\includegraphics[width=0.48\textwidth]{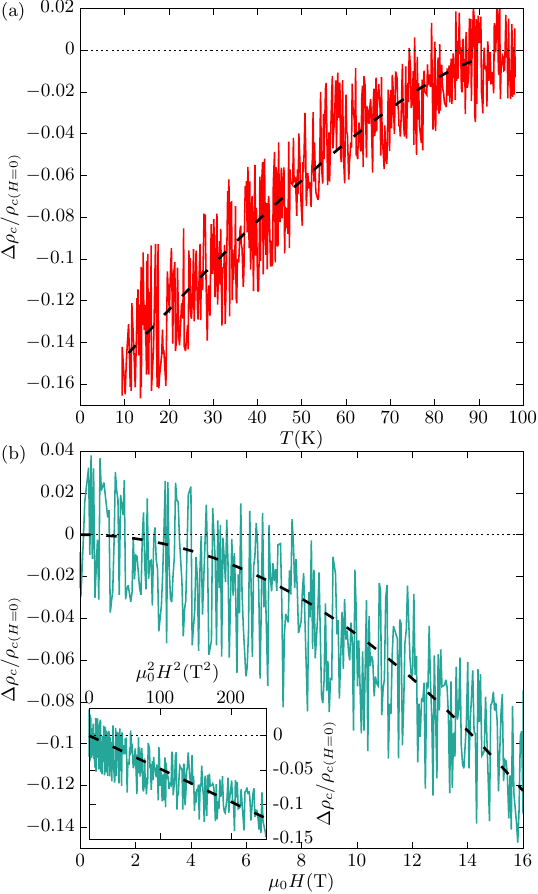}
\end{center}
	\caption{(a) Temperature dependence of the magnetoresistance with $\mu_0 H$=16\,T. (b) Field dependence of the magnetoresistance for $T$=16.15(5)\,K. Inset: Magnetoresistance as a function of $H^2$ for $T$=16.15(5)\,K. In all cases, $H$$\parallel$$I$$\parallel c$.  The dashed lines represent fits using equation \ref{delta-r_fit}. }\label{mrvst}
\end{figure}

Figures \ref{mrvst}(a) and \ref{mrvst}(b) illustrate how the longitudinal magnetoresistance varies with temperature for $\mu_0 H$=16\,T and with the magnetic field strength for $T$=16.15(5)\,K, respectively. 
Similar observations have been reported in Ba(Fe$_{1-x}$Co$_x$)$_2$As$_2$ and LiFeAs for the longitudinal magnetoresistance in the $ab$-plane\cite{rullier2013longitudinal}.

\section{Discussion}
\label{discussion}

Negative longitudinal magnetoresistance cannot be explained solely within a semiclassical description of multiband models that accounts for only the effect of the magnetic field through the semiclassical Lorentz force, as this force  vanishes when $H$$\parallel$$I$. 
A natural origin of NLMR lies in the scattering from spin fluctuations  \cite{PhysRevLett.20.665,doi:10.1143/JPSJ.44.122,doi:10.1143/JPSJ.45.466,PhysRevB.51.9253,majumdar1998dependence}, the spectrum of which can potentially be affected by the external magnetic field.

In this context, it is useful to consider prior studies exploring the spectrum of spin fluctuations in FeSe as a function of temperature. 
Different techniques sensitive to the presence of spin fluctuations, such as  nuclear magnetic resonance (NMR) \cite{baek2015orbital,PhysRevLett.114.027001,doi:10.7566/JPSJ.87.013704}, inelastic neutron scattering (INS) \cite{PhysRevB.91.180501,wang2015strong,wang2016magnetic,ma2017prominent,chen2019anisotropic} or resonant inelastic x-ray scattering (RIXS) \cite {PhysRevB.99.014505} have been applied to the  case of FeSe.
One important early observation is that even if long-range magnetic order is absent, FeSe does present strong stripe spin fluctuations \cite{PhysRevB.91.180501,wang2015strong}.
A key question is whether the magnetic response is appreciable above $T_s$ and how it evolves upon cooling below $T_s$.
Concerning the first point, a dependence on the energy scale of the probed spin fluctuations has been noticed \cite{vaknin2016magnetic,baek2015orbital,wang2015strong}. NMR experiments probing a smaller energy scale tend to evidence a spin-fluctuation contribution only below $T_s$, while INS experiments detect their presence already above $T_s$ and note a significant enhancement upon entering the nematic phase.
More recently, the importance of spin fluctuations in the nematic phase has been considered to interpret anisotropies of the in-plane magnetostriction \cite{PhysRevB.97.104107}, of the in-plane resistivity \cite{PhysRevLett.117.127001} and of the optical response \cite{PhysRevB.98.094506}. 
Furthermore, in FeSe$_{1-x}$S$_x$, a connection between the linear in $T$ resistivity with spin fluctuations was established at low temperature inside the nematic phase \cite{PhysRevResearch.2.013309}.
Given the different results which point to an enhancement of the spin fluctuations in the nematic phase, it is indeed plausibly that in such fluctuations lies the origin of the observed NLMR.

The simplest models that describe NLMR due to spin fluctuations trace back to the seminal papers \cite{doi:10.1143/JPSJ.44.122,doi:10.1143/JPSJ.45.466}. The model studied in \cite{doi:10.1143/JPSJ.45.466} corresponds to an antiferromagnet where electrons that form an $s$-like band scatter with localized $d$ electrons. The dynamical susceptibility of the system, $\chi_\mathbf{q}(\omega)$, has peaks at $\mathbf{q}=0$ and at $\mathbf{q}=\mathbf{Q}$, the antiferromagnetic wave vector. 
The contribution to the resistivity from the exchange of momentum $\mathbf{Q}$ with spin fluctuations is found to yield a magnetoresistivity
\begin{equation}
	\Delta \rho(H,T) = \rho_{(H)}-\rho_{(H=0)} = - a(T) T^{\alpha} H^2.
\label{delta-r}
\end{equation}
Here, the exponent $\alpha$ can vary depending on the dimensionality, and the function $a(T)$ is related to the dynamical spin susceptibility. 
The external magnetic field promotes an enhancement of the  uniform $\mathbf{q}=0$ mode at the expense of the staggered $\mathbf{q}=\mathbf{Q}$ component. The manner in which this interplay affects the resistivity is contained in  $a(T)$. 

Certainly, the rich interplay between electronic, lattice, orbital, and spin degrees of freedom makes the situation in FeSe far more complex than what is captured by the simple $s$-$d$ model. In addition to the complexity arising from its multiorbital nature, the conducting electrons in FeSe themselves act as a source of spin fluctuations. Moreover, while there are well-established models for the electronic structure of FeSe in the tetragonal phase \cite{rhodes2017strongly}, the description of the band structure in the nematic phase as a function of temperature remains less settled \cite{rhodes2022fese}. Although a comprehensive understanding of the various ingredients shaping the electronic system in FeSe is likely crucial for a quantitative assessment of the observed negative longitudinal magnetoresistance, the $s$-$d$ model provides a foundational framework for a qualitative understanding of the phenomenon. It is in particular valuable because it accounts for the effect of the magnetic field on the scattering with antiferromagnetic fluctuations, which as explained above, are known to be important in FeSe.
This perspective has also been previously applied to other iron-based compounds such as Ba(Fe$_{1-x}$Co$_x$)$_2$As$_2$ and LiFeAs \cite{rullier2013longitudinal}.

In Ba(Fe$_{1-x}$Co$_x$)$_2$As$_2$, for $x<0.06$, there is long-range antiferromagnetic order and in the paramagnetic phase, the observed NLMR can be described by the equation $\Delta \rho(H,T) \propto H^2 T/(T+\Theta)$, with $\Theta$ being the Néel temperature. 
For larger values of $x$, it is observed that  $\Delta \rho(H,T) \propto H^2 T/ (T+\Theta)^2$, with $\Theta \leq 0$ depending on $x$.
The same applies to LiFeAs with $\Theta=0$, which is an interesting case since like FeSe, LiFeAs does not present long-range magnetic order; however, unlike FeSe, it preserves the tetragonal phase across the entire temperature range. 
These examples show that the temperature dependency of the NLMR, expressed in the function $a(T)$  in equation \ref{delta-r},  can strongly depend on the nature of the spin fluctuations in a given system. 

In the case of FeSe, based on INS experiments it is reasonably to anticipate a dependency on $T_s-T$ within a certain temperature range. Indeed, close to $T_s$, a power law behavior of the dynamical susceptibility has been reported \cite{wang2015strong}. We find that for different samples, the observed NLMR can be effectively fitted with
\begin{equation}
	\frac{\Delta \rho(H,T)}{\rho(H=0,T)} = [\Delta_0 + \nu (T_s-T)^{\beta}] H^2.
\label{delta-r_fit}
\end{equation}
Here $\beta= 1.3(4)$, $\nu=-1.9(5)\times 10^{-6}$\,T$^{-2}$K$^{-1.3}$ and $\Delta_0=-1.8(5)\times 10^{-5}$\,T$^{-2}$. The latter parameter accounts for a very small NLMR above $T_s$ which, e.g. at 110\,K, is barely measurable. 

Certainly, the \( ab \)-plane magnetotransport properties of FeSe exhibit a complex temperature evolution within the nematic phase and it is possible that the here observed NLMR may also display a different temperature dependence at lower temperatures. In particular, a characteristic temperature \( T^* \sim 20 \) K has been identified, where both the Hall coefficient and longitudinal resistivity undergo significant changes.
In Ref.  \cite{kasahara2016giant},  \( T^* \) is defined as the temperature at which the derivative of the resistivity with respect to temperature reaches a minimum in zero magnetic field. We note that the signal-to-noise ratio in our \( c \)-axis resistivity data limits a detailed analysis of its derivative as a function of temperature; thus, we cannot confirm a similar crossover of the $c$-axis longitudinal magnetoresistance at \( T^* \), but we do not exclude this possibility.
In Ref.  \cite{PhysRevResearch.2.013309}, \( T^* \) is defined as the temperature at which the resistivity reaches a maximum in the presence of a strong magnetic field (34\,T), which is beyond the field range of our experiments. Extending our results to such high-field regimes would certainly be of interest.
In any case, the presence of NLMR is robust and, in line with \cite{rullier2013longitudinal}, it can also be qualitatively understood in agreement with INS experiments, based on the effects of a magnetic field on the scattering with spin fluctuations.

Lastly, we would like to note that a negative contribution to the magnetoresistivity might also be present in other field and current configurations. The fact that negative magnetoresistivity is observed exclusively when $H$$\parallel$$I$$\parallel$$c$ could be attributed to two main factors. 
Firstly, spin fluctuations in FeSe have been reported to be anisotropic and 
stronger along the $c$-axis \cite{ma2017prominent}. 
Therefore, the effects induced by a magnetic field could be more pronounced when applied in this direction. 
Secondly, in all the configurations exhibiting transverse magnetoresistance, there is a positive contribution arising semiclasically from the Lorentz force. This effect could dominate, resulting in $\Delta \rho \mathord / \rho_{(H\mathord=0)} \mathord > 0$.

\section{Conclusion}
\label{conclusions}

We have studied the $c$-axis resistivity of FeSe in single crystals.
Our main finding is the emergence of negative longitudinal magnetoresistance, which is markedly enhanced below $T_s$.
Negative longitudinal magnetoresistance cannot arise from the Lorentz force and is often attributed to spin fluctuations. These fluctuations provide a scattering mechanism for electronic resistivity and diminish upon increasing the magnetic field. Our results align with this interpretation and underscore the importance of spin fluctuations in the magnetotransport properties of FeSe, within its nematic phase.

\section{Acknowledgements}
M.L.A and G.N. acknowledge financial support from U. N. de Cuyo 06/C018 T1. 
M.L.A and J.I.F. acknowledge support from the Alexander von Humboldt Foundation. 
G.N. acknowledges financial support from ANPCYT PICT2018-1533.

\appendix
\section{} 
\label{ApenA}

\begin{figure}[h]
\begin{center}
\includegraphics[width=0.48\textwidth]{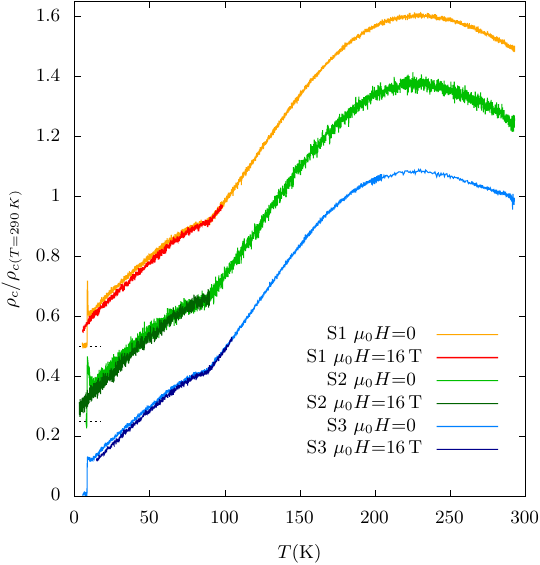}
\end{center}
\caption{Temperature dependence of the normalized magnetoresistance for $\mu_0 H$$\parallel$$c$=0 and 16\,T. The curves for each sample are vertically shifted with the same factor.}\label{Ap1}
\end{figure}

In the main text, we presented data for one sample. Here, we include additional data for two other samples exhibiting the same phenomenology.
Figure \ref{Ap1} shows the normalized $c$-axis resistivity for three FeSe samples under a magnetic field of $\mu_0 H$$\parallel$$c$=0 and 16\,T. The resistivity maxima are at $T_{c}^{max}$= 229(1), 228(1) and 227(1)\,K for samples S1, S2 and S3, respectively. In all three samples, the NLMR is clearly observed below $T_s$. Sample S1 is the one presented in the main part of this work.

\begin{figure}[h!]
	\begin{center}\vspace{3mm}
\includegraphics[width=0.43\textwidth]{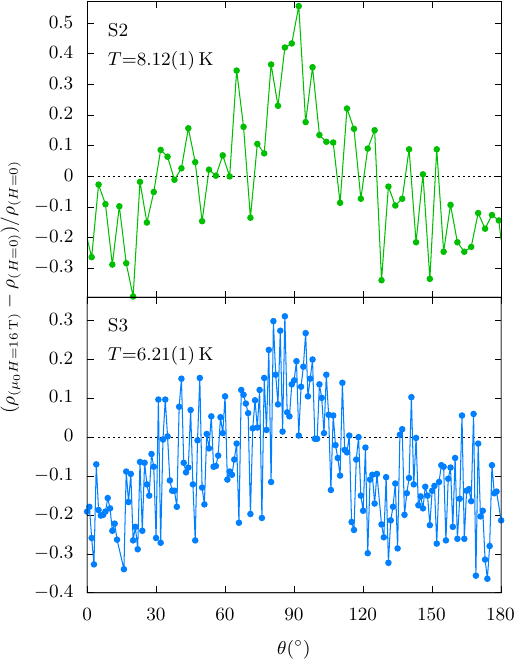}
\end{center}
\caption{Angular dependence of the magnetoresistance for $\mu_0 H$=16\,T observed in samples S2 and S3 at temperatures 8.12(1) and 6.21(1)\,K respectively. The angle is measured between the magnetic field and the $c$-axis of the sample, where $\theta$=0 denotes $H$ parallel to the $c$-axis.}\label{Ap2}
\end{figure}
Additionally, figure \ref{Ap2} presents the angular dependence for the current along the $c$-axis in samples S2 and S3 at various temperatures. Here, similarly to figure \ref{rvsth} in the main text, the magnetoresistance is negative when the magnetic field is parallel to the $c$-axis and positive when the field is oriented within the $ab$-plane.

\begin{figure}[t]
\begin{center}
\includegraphics[width=0.44\textwidth]{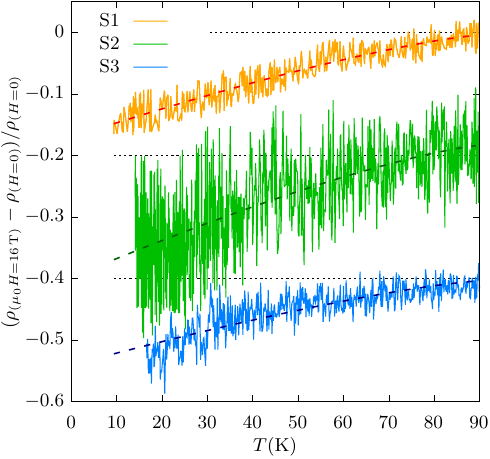}
\end{center}
\caption{Temperature dependence of the magnetoresistance for $\mu_0 H$$\parallel$$c$=16\,T and the fit of equation \ref{delta-r_fit} in dashed line. The curves for each sample are vertically shifted by the same factor for clarity.}\label{Ap3}
\end{figure}
Finally, figure \ref{Ap3} presents the magnetoresistivity for $\mu_0$$H$$\parallel$$c$=16\,T and the fit of equation \ref{delta-r_fit} for samples S1, S2 and S3. The parameters obtained from the fit are: $\beta_{S2}$=1.3(4), $\nu_{S2}$=$-$2.4(7)$\times$$10^{-6}$\,T$^{-2}$K$^{-1.3}$, $\Delta_{0 S2}$=6(1)$\times$$10^{-6}$\,T$^{-2}$, $\beta_{S3}$=1.3(4), $\nu_{S3}=-1.5(5)\times 10^{-6}$\,T$^{-2}$K$^{-1.3}$ and $\Delta_{0 S3}$=$-$1.5(5)$\times$$10^{-5}$\,T$^{-2}$. The parameters for S1 are in Section \ref{results} of the main text.

\break

\section{} 
\label{ApenB}

\begin{figure}[t]
\begin{center}
\includegraphics[width=0.48\textwidth]{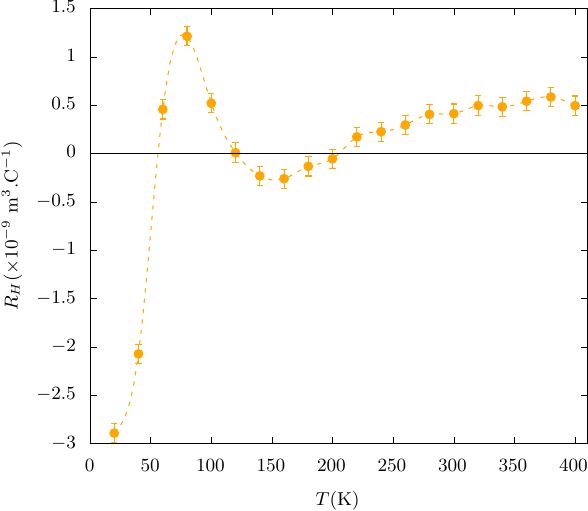}
\end{center}
	\caption{Temperature dependence of the Hall coefficient. The dashed line is a guide to the eye.}\label{RH}
\end{figure}

Figure \ref{RH} presents the temperature dependence of the Hall coefficient for FeSe, obtained as d$\rho_{xy}^{ab}$/d$H$ for $H\parallel c$ in the linear response regime. Similar behavior has been reported in the literature \cite{kasahara2014field, PhysRevLett.115.027006}.

\bibliography{mibib}

\end{document}